\documentclass[conference]{IEEEtran}
\IEEEoverridecommandlockouts
\usepackage{hyperref}
\usepackage{algorithm}
\usepackage{physics}
\usepackage{cite}
\usepackage{amsmath,amssymb,amsfonts}
\usepackage{algorithmic}
\usepackage{graphicx}
\usepackage{textcomp}
\usepackage{xcolor}
\usepackage{braket}

\def\BibTeX{{\rm B\kern-.05em{\sc i\kern-.025em b}\kern-.08em
    T\kern-.1667em\lower.7ex\hbox{E}\kern-.125emX}}
\begin{document}

\title{A Quantum Approach to solve N-Queens Problem}

 %\author{
 %  Santhosh G S\\
 %  \affil{Sri Sivasubramaniya Nadar College of Engineering, Rajiv Gandhi Salai (OMR), Kalavakkam, 603110, Tamil Nadu, India}
%\texttt{first1.last1@xxxxx.com}
%  \and
%   Joshi, Piyush\\
%    \affil{Indian Institute of Space Science and Technology, Valiamala, Thiruvananthapuram, 695547, Kerala, India}
%    \and
%     Barui, Ayan\\
%    \affil{ Indian Institute of Science Education and Research Kolkata, Mohanpur, 741246, West Bengal, India}
%    \and 
%    K. Panigrahi, Prasanta\\
%    \affil{ Indian Institute of Science Education and Research Kolkata, Mohanpur, 741246, West Bengal, India}
  
 % \texttt{first2.last2@xxxxx.com}
%}

 \author{
     \IEEEauthorblockN{
         Santhosh 
         G S \IEEEauthorrefmark{1}, 
         Piyush 
    Joshi\IEEEauthorrefmark{2},
        Ayan
        Barui\IEEEauthorrefmark{3}, and 
         Prasanta K. Panigrahi\IEEEauthorrefmark{3,*}
     }
      \IEEEauthorblockA{
         \IEEEauthorrefmark{1} Sri Sivasubramaniya Nadar College of Engineering,Rajiv Gandhi Salai (OMR), Kalavakkam, 603110, Tamil Nadu, India
         }
     \IEEEauthorblockA{
         \IEEEauthorrefmark{2} Indian Institute of Space Science and Technology, Valiamala, Thiruvananthapuram, 695547, Kerala, India
     }
     
     \IEEEauthorblockA{
         \IEEEauthorrefmark{3} Indian Institute of Science Education and Research Kolkata, Mohanpur, 741246, West Bengal, India}
     }
\maketitle
\begin{abstract}
In this work, we have introduced two innovative quantum algorithms: the Direct Column Algorithm and the Quantum Backtracking Algorithm to solve N-Queens problem, which involves the arrangement of $N$ queens on an $N \times N$ chessboard such that they are not under attack from each other on the same row, column and diagonal. These algorithms utilizes Controlled W-states and dynamic circuits, to efficiently address this NP-Complete computational problem. The Direct Column Algorithm strategically reduces the search space, simplifying the solution process, even with exponential circuit complexity as the problem size grows, while Quantum Backtracking Algorithm emulates classical backtracking techniques within a quantum framework which allows the possibility of solving complex problems like satellite communication, routing and VLSI testing.
\end{abstract}


\section{Introduction}
 The field of computer science presents various computational problems, which encompass tasks, questions, or challenges that demand computer-based or algorithmic solutions. These problems exhibit a wide spectrum of complexity, scope, and applications, making them fundamental in the realm of computer science \cite{brookshear1991computer}. Computational problems are often categorized into different complexity classes, based on the computational resources required for their efficient solutions. One prominent class of problems is NP-Complete (Non-Deterministic Polynomial Complete) problems. In this context, 'P' stands for Polynomial time complexity, indicating that problems in class 'P' can be solved in polynomial time. Conversely, 'NP' stands for Non-deterministic Polynomial, and it represents a class of problems for which a proposed solution can be verified in polynomial time. However, it remains an open question whether all problems in 'NP' can also be solved in polynomial time (i.e., whether P equals NP).

 Consider NP-Complete problems, which form a subset of the 'NP' class. NP-Complete problems share a common characteristic: if a single problem from this class is solved, it implies that all problems within this category can be considered solved. These problems have a rich history in the field of computer science, posing significant challenges and opportunities for exploration. In this paper, we will focus on our efforts to address a well-known NP-Complete problem, the notorious N-Queens problem \cite{erbas1992different}. We will detail our approach and findings related to this intriguing challenge.

The paper's structure is outlined as follows: Section \ref{s2} defines and elaborates on the intricacies of the N-Queens problem. In Section \ref{s3}, we briefly explain the approach and the solution presented in the previous paper. Section \ref{s4} encompasses the novel improvements that have been integrated into the previous work. In Section \ref{s5}, we explore two novel algorithms, each employing varied strategies to tackle this problem. Finally, we conclude the paper by summarizing our findings and presenting future discussion points.

\section{N-Queens Problem}
\label{s2}
The N-Queens problem is a classic example of an NP-Complete problem, known for its challenging nature. Since the 1850s, first emerging as a mathematical recreation, the problem has seen numerous published attempts \cite{bell2009survey}.  Pauls first established the existence of at least one solution for the generalized problem when $N>3$ and multiple proven bounds suggest the number of solutions increasing exponentially with increasing $N$ \cite{pauls1874maximalproblem}.  On the other hand, the problem is well-suited for search-based algorithms. For instance,  the backtracking search algorithm \cite{bitner1975backtrack}, which employs a recursive approach, can generate all potential solution sets for an $N \times N$ board. Imagine a world where the chessboard's size is not limited to the traditional $8\times8$ squares; it can be of any dimension as long as it remains a square. We will refer to it generally as the $N\times N$ Chessboard. In this world, the only chess pieces available are Queens, meaning that Queens are the sole occupants of the chessboard. With this setup in mind, let us consider a fundamental question: How many Queens can be placed on an $N\times N$ Chessboard, such that none of the Queens attack each other? The answer is N Queens. In an $N\times N$ chessboard, a maximum of N Queens can be placed in such a way that they do not threaten each other. However, not all arrangements of Queens satisfy this condition. Only specific arrangements allow all the Queens to coexist harmoniously. This problem, known as the N-Queens problem, challenges us to find all the valid combinations of placing N Queens on an $N\times N$ chessboard.  In practice, however, this algorithm faces challenges in finding substantially distinct solutions within the solution space due to its high time complexity of $\mathcal{O}(N!)$ \cite{sosic1994efficient}. This complexity is what continues to make the N-Queens problem a subject of study and exploration \cite{stone1987efficient}. From Table \ref{t1}, it becomes evident that there is no straightforward way to model the relationship between the $N\times N$ chessboard and the corresponding number of solutions. In the case of the N-Queens Problem, it has been demonstrated that finding all the solutions is beyond the \#P-class \cite{hsiang2004hardness}.  This, in turn, confirms the absence of a closed-form solution for the problem and motivates computer searches to be particularly helpful. However, classical computational methods are very limited in addressing such challenges due to the need to process large datasets and explore extensive solution spaces. In recent years, quantum technology has shown rapid advancements to approach computational quantum advantage for different numerical tasks \cite{mcclean2016theory,peruzzo2014variational,biamonte2017quantum}, particularly optimization problems (demonstrated recently even on near-term devices)

\begin{table}[h]
\centering
\caption{Number of Solutions for the N-Queens Problem\\
The following table was referred from TABLE 1 of the article \cite{10.1145/131214.131227}}
\label{t1}
\begin{tabular}{|c|c|}
\hline
$N$ & Number of Solutions \\
\hline
1 & 1 \\
2 & 0 \\
3 & 0 \\
4 & 2 \\
5 & 10 \\
6 & 4 \\
7 & 40 \\
8 & 92 \\
%9 & 352 \\
%10 & 724 \\
%11 & 2680 \\
%12 & 14200 \\
%13 & 73732 \\
%14 & 365596 \\
%15 & 2279184 \\
%16 & 14772512 \\
%17 & 95815104 \\
%18 & 666090624 \\
\hline
\end{tabular}
\end{table}

Until today, numerous algorithms have been developed to solve the N-Queens problem, each with varying time complexities. While we won't delve into every algorithm, let's briefly explore a few notable ones. The first algorithm is the Backtracking algorithm, which has a worst-case time complexity of $O(N!)$ \cite{thada2014performance}. It is a classic approach for solving the N-Queens problem. The next algorithm, introduced by Sosic and Gu in 2010, is a linear-time algorithm. It employs a pattern-based approach to produce at least one unique solution for all values of N greater than 3. The time complexity of the Sosic and Gu algorithm is $O(N)$, a significant improvement over the Backtracking algorithm's time complexity \cite{sosic1994efficient}. Over the years, many efforts have been made to solve the N-Queens problem in the quantum regime. A quantum-inspired differential evolution algorithm for solving the N-Queens problem was proposed by Draa et al. \cite{draa2011quantum}. Recent developments include the work of Codognet where the QUBO model was used to solve the N-Queens problem on the D-Wave quantum annealer \cite{codognet2023encoding}. 

We might wonder why there is such persistent effort to solve the challenging N-Queens problem. The reason lies in the diverse applications of the N-Queens problem and its variations across various domains. These applications encompass not only neural networks \cite{che2018recurrent} algorithm benchmarking \cite{gent2017complexity} and parallel memory storage schemes \cite{erbas1992storage} but also extend to fields like VLSI (Very-Large-Scale Integration)  design and testing \cite{williams1986vlsi}, AI and robotics \cite{10.1007/3-540-36127-8_13}, and more. The versatility of the N-Queens problem makes it a valuable resource for addressing complex real-world challenges and advancing the state of the art in numerous areas of science and technology.

\section{Previous Work}
\label{s3}
\begin{figure}[h]
    \centering
    \includegraphics[scale=0.45]{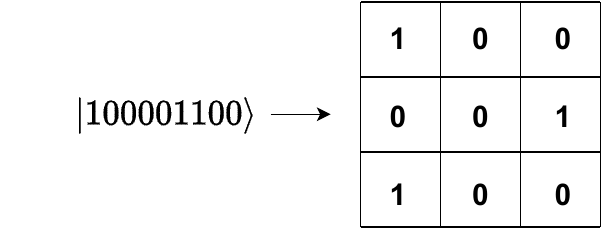}
    \caption{$\ket{1}$ and $\ket{0}$ formulated in the braket notation which is realised in the form of a chessboard}
    \label{represen}
\end{figure}

Our solution builds upon the foundations laid by the work of Jha et al\cite{jha2018novel}, and we would like to acknowledge and credit this work throughout our paper. We will use notations and representations introduced in \cite{jha2018novel}. Additionally, we will introduce two new solutions inspired by the ideas presented in \cite{jha2018novel}. The authors of this work leveraged quantum properties, such as superposition and entanglement, to tackle the N-Queens problem. In their approach, they proposed a novel way of representing the positions of Queens on a chessboard using an $N\times N$ matrix. The squares are numbered from 1 in the top-left corner to N squared in the bottom-right corner, with numbering proceeding from left to right and top to bottom. This concept can be translated into a Quantum Circuit comprising $N\times N$ qubits, where each qubit is assigned a number from 1 to $N\times N$. If a specific position on the chessboard is occupied by a Queen, the corresponding qubit is encoded in an excited state $\ket{1}$. All other qubits remain in the ground state $\ket{0}$. This is shown in Figure \ref{represen}.

Let's discuss their approach to solving the N-Queens problem.

They decomposed this complex problem into three simpler criteria. A set of Queen positions satisfies all three criteria together, making it a potential solution to the N-Queens problem.

\textbf{Row Criteria:} The row criteria is satisfied when there is exactly one queen at any position in every row of the chessboard. In quantum circuit terms, this means that exactly one qubit in every row should be in the excited state $\ket{1}$.

\textbf{Column Criteria:} The column criteria is met when there is exactly one queen at any position in every column of the chessboard. In the quantum circuit, this translates to exactly one qubit in every column being in the excited state $\ket{1}$.

\textbf{Diagonal Criteria:} The diagonal criteria present the most challenging aspect of the problem. The chessboard contains a total of $(4N-2)$ diagonals. On each of these diagonals, there should be at most one queen, and it's acceptable for many diagonals to have no queens at all. In quantum circuit terms, this implies that on every diagonal, there should be a maximum of one qubit in the excited state $\ket{1}$. Notably, a considerable number of diagonals might not have any qubit in the excited state $\ket{1}$ at any of their positions. Let us now delve into how the authors of \cite{jha2018novel} have addressed these criteria in detail in the following subsections:

\subsection{Solving Row Criteria}

We begin by exploring all the possible setups for positioning queens on a chessboard, regardless of the number of queens. Each square on the chessboard can either have a queen placed on it or remain empty, resulting in two potential options for each square. Given that there are $N \times N$ squares on an $N \times N$ chessboard, there's a possibility of having $2^{N^2}$ combinations for $N$ queen placements.
\begin{equation} \label{e1}
 \ket{00..0} \xrightarrow[]{H^{\otimes N^2}} \frac{1}{\sqrt{2^{N^2}}}\sum_{x=0}^{2^{N^2}-1} \ket{x}
\end{equation}

Let us introduce a condition: on an $N \times N$ chessboard, we aim to place exactly $N$ queens. With this requirement in mind, the search scope narrows down to a combinatorial challenge with ${N \times N \choose N}$ potential combinations. In a clever approach, the authors have devised a method to configure the quantum circuit, allowing the outcomes of measurements to span all the potential combinations within the search domain, satisfying the condition of rows without conflicts. For each group of $N$ qubits, representing a single row on the chessboard, the authors have introduced the concept of implementing what they refer to as an $N$-Qubit W-State, represented in Eq. (\ref{wn}).
\begin{equation}
\label{wn}
    \ket{W_N}=\frac{1}{\sqrt{N}}(\ket{100\dots0} + \ket{010\dots0} + \dots + \ket{000\dots1})
\end{equation}
\begin{equation}\label{R}
\ket{R}=\ket{W_N} = \frac{1}{\sqrt{N}} \left [ \ket{1} + \ket{2} +..... \ket{N}  \right ] = \frac{1}{\sqrt{N}}  \sum_{i=1}^{N} \ket{x_i}
\end{equation}
where $\ket{x} \in \ket{100\dots0}$ to $\ket{000\dots1}$.
Here, each vector $\ket{x_i}$ $\left ( i \in \left \{1,2,... N \right \}\right )$ represents all possible configurations of a particular row. When this process is applied to every row on the chessboard, it effectively ensures that the arrangement adheres to the row criterion without any conflicts. Taking $\ket{R_j}$ as the state representing the $j$th row configurations, the tensor product of N of these states, represented as $\ket{\Psi_R}$ is as follows: 

\begin{equation}\label{eq:3}
    \ket{\Psi_R} = \ket{R_1} \otimes \ket{R_2} \otimes .... \ket{R_N}  
\end{equation}
where $\ket{\Psi_R}$ represents all the possible combinations  of $N^2$ qubit vectors, each of which fulfills the \textbf{Row criterion}. This reduces the subspace, taking it from the general subspace of $2^{N^2}$ to $N^N$. However, it's essential to note that the reference paper \cite{jha2018novel} doesn't provide a precise step-by-step procedure for generating a W-State. Instead, the authors illustrate the algorithm using a $4 \times 4$ chessboard as an example. In this particular case, they employ a specialized technique to craft a W-State tailored to 4 qubits. It's crucial to understand that this technique cannot be universally extended to accommodate any arbitrary number of qubits. Despite this limitation, the algorithm retains its appeal due to its unique ability to set up the quantum circuit directly in a way that ensures the row criteria are met without the need for complicated post-processing. 

\subsection{Solving Column Criteria}

\begin{figure}[h]
    \centering
    \includegraphics[width=0.45\textwidth]{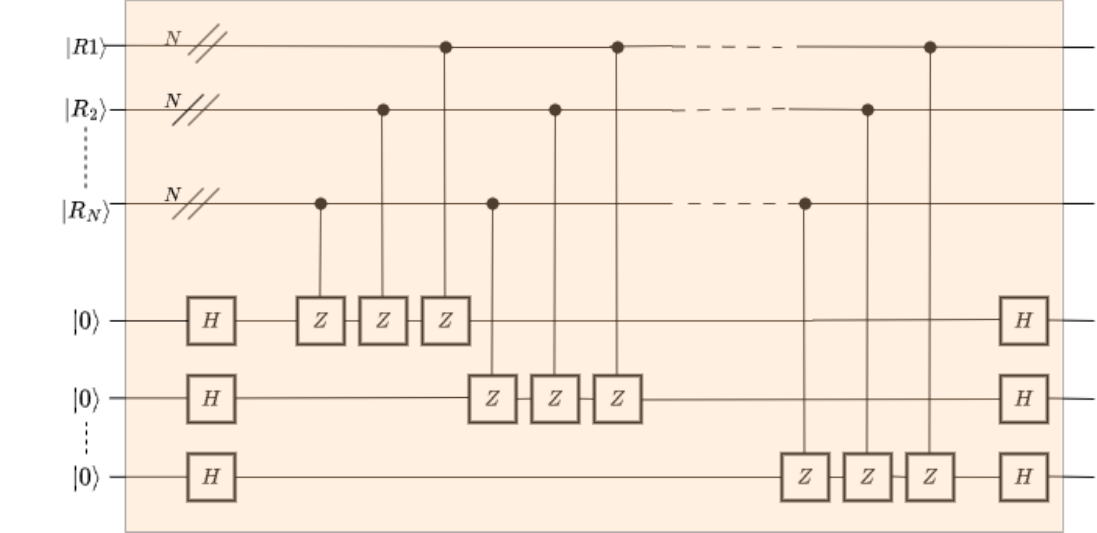}
    \caption{Quantum circuit design for solving the \textbf{column criteria}. Here $\ket{R}$ is the configuration of each row obtained by the W State as shown n Eq. \ref{R}.}
    \label{cc}
\end{figure}

After successfully getting all the configurations of placing $N$ queens in $N \times N$ chessboard stored in $\Psi_R$ that fulfills the row criteria, our exploration is now limited to $N^N$ combinations. To handle and enhance the column criteria, a group of $(N-1)$ ancillary qubits is introduced. These qubits act as indicators, signaling that if all these ancillary qubits are in the excited state $\ket{1}$, the specific combination adheres to the column criterion. Let's point out the process step by step:

\begin{enumerate}
\item Apply a Hadamard ($H$) gate to each ancillary qubit. This transforms the ancillary qubits from their original $Z$-basis to the $X$-basis.
\item Consider qubit $q_{i,j}$ which represents the quantum state $\ket{0}$ or $\ket{1}$ belonging to the $i$-th row and $j$-th column. For each $j$-th qubit present in all the rows, apply a controlled phase $(CP)$ gate controlled by the $j$-th qubit and applied to $j$-th ancillary qubit. Repeat this process iteratively until $j$ reaches $(N-1)$, excluding the last qubit of every row. This process is depicted in Figure \ref{cc}.
\item It's implicit that if every one of the $(N-1)$ columns contains just a single qubit in the excited state $\ket{1}$, the ultimate qubit naturally resides in the last column.
\item Implement another Hadamard ($H$) gate operation on all the ancillary qubits. The outcome of this step ensures that if a column contains multiple qubits in the excited state $\ket{1}$ or none at all, the associated ancillary qubit remains in the ground state $\ket{0}$.
\end{enumerate}

It does not matter if there are multiple qubits in the same column because only if all the ancillary qubits are in the excited state $\ket{1}$, the column criteria is fulfilled. In effect, only when both the row and column criteria are met, all the ancillary qubits transition to the excited state $\ket{1}$. This signifies the fulfillment of the required conditions. In this manner, the quantum circuit proficiently examines the interplay between the row and column criteria, pinpointing valid combinations that satisfy both sets of requirements and efficiently navigating the complex constraints of the $N$-Queens problem.

\subsection{Solving Diagonal Criteria}

After successfully fulfilling both the row and column criteria, our search space is further narrowed down to $N!$ distinct states. For an $N\times N$ chessboard, the process involves using $(N-1)$ ancillary qubits to verify column criteria and $\frac{N(N-1)}{2}$ ancillary qubits for diagonal criteria. The central concept here revolves around scrutinizing every possible pair of rows to identify the presence of excited qubits diagonally. We need to include only $\frac{N(N-1)}{2}$ ancillary qubits, because there are fewer number of diagonal checks required as the row and column criteria are already satisfied. 

Here's how the process unfolds:

\begin{enumerate}
\item Initialize each ancillary qubit to the excited state $\ket{1}$.
\item The diagonal checks are executed using Toffoli gates, with each diagonal check corresponding to one Toffoli gate. For every pair of rows, select the relevant ancillary qubit associated with that pair.
\item For each pair of rows, identify the pairs of qubits situated diagonally from the chosen rows. Apply a Toffoli gate to the ancillary qubit which represents the corresponding diagonal which is being checked, controlled by these two pairs of diagonal qubits. Repeat this procedure for all pairs of diagonal qubits stemming from the selected rows.
\item Iterate through every pair of rows, performing the same Toffoli gate operation on the corresponding ancillary qubit representing the current diagonal under check. If any diagonally placed pair of qubits is excited, it will transition the related ancillary qubit back to the ground state $\ket{0}$.
\end{enumerate}

Following these operations, if all the ancillary qubits designated for diagonal criteria remain in the excited state $\ket{1}$, then the states that meet this condition are the ones that satisfy the diagonal criteria. In summary, if both the $(N-1)$ ancillary qubits for column criteria and the $\frac{N(N-1)}{2}$ ancillary qubits for diagonal criteria remain in the excited state $\ket{1}$, the resulting states—excluding the ancillary qubits themselves represent the solutions to the $N$-Queens problem. These states signify the valid positions for placing $N$ queens on an $N\times N$ chessboard, ensuring that no two queens threaten each other.
\section{Improvements to the Previous Solution}
\label{s4}
In this work we have improved upon the complexity of the existing algorithm by including efficient preparation of $W_N$-states and introducing the use of dynamic circuits. 

\subsection{Generalized $W_N$ States}

The concept of using $W_N$ states to initialize a quantum circuit in order to satisfy the column criteria was previously introduced. This was exemplified with a $4\times 4$ chessboard, which necessitates the creation of a 4-qubit $W$-state, a relatively straightforward task. However, the methods employed for preparing 4-qubit $W$-states are not easily adaptable for generating $W$-states for varying numbers of qubits. Subsequently, it was realized that creating $W$-states for $2K$ qubits, where $K$ is a whole number, is comparatively simpler than devising methods for arbitrary whole numbers. As a solution, a more comprehensive approach was discovered in Cruz's work \cite{cruz2019efficient}, detailing a generalized method for preparing $N$-qubit $W$-states, where $N$ can be any whole number. To make the algorithm more practical and extensible, the aforementioned method for generating $W$-states for any number of qubits was integrated.
\begin{figure}
    \centering
    \includegraphics[width=0.35\textwidth]{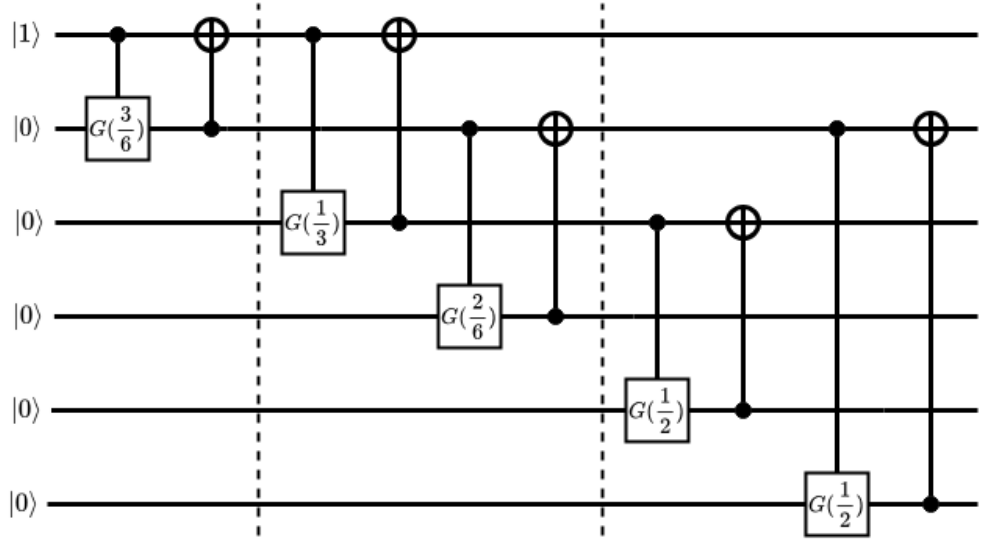}
    \caption{Logarithmic Complexity Circuit for $\ket{W_6}$. The following circuit design was referred from the article \cite{cruz2019efficient}}
    \label{fig: Circuit illustrating logartmic time complexity}
\end{figure}
Furthermore, there was an advancement in the process of filtering states that satisfy both row and column criteria. In the previous paper, this involved transitioning the ancillary qubits from the $Z$-axis to align with the $X$-axis, then executing controlled phase flip operations on these ancillary qubits, controlled by the qubits of the corresponding column. The ancillary qubits would then be switched back to the $Z$-axis configuration. This improvement in the approach to satisfying the column criteria refined the algorithm and enhanced its practical applicability.

\subsection{Removing the Phase operation from the Column criteria}

An alternative approach involves using a controlled bit flip gate instead of a controlled phase flip gate. This change eliminates the need to switch between the $Y$-axis and $Z$-axis configurations. This alteration decreases the number of gates by 2 for each ancillary qubit dedicated to column criteria. Additionally, the controlled bit flip gate is simpler to implement and offers improved noise reduction in real quantum computers. In the Qiskit framework, we implemented a controlled bit flip gate using the CX-gate (CNOT gate), and a controlled phase flip gate using the CZ-gate.

\subsection{Introducing Dynamic Circuits}

When the previous paper was proposed, dynamic circuit features were not yet introduced in any of the quantum simulation frameworks. However, today, almost every quantum computing simulation framework supports dynamic circuits. They offer a powerful way to leverage quantum computing capabilities for various applications, allowing exploration of quantum states and behaviors that might not be achievable using static circuits. However, working with dynamic circuits does present challenges, including efficient parameter optimization and managing complexity as circuits become more adaptive. Table \ref{t2} presents a summary of the introduction years and versions for various quantum computing frameworks that support dynamic circuits:

\begin{table}[h]
\centering
\caption{Qiskit Versions for dynamic circuits}
\label{t2}

\renewcommand{\arraystretch}{1.5}
\begin{tabular}{|c|c|c|}
\hline
\textbf{Framework} & \textbf{Year Introduced} & \textbf{Version} \\
\hline
Qiskit & 2020 & 0.23.0 \\
OpenQASM & 2021 & 3.0.0 \\
Cirq & 2021 & 1.9.0 \\
%ProjectQ & 2022 & 0.25.0 \\
%Strawberry Fields & 2023 & 0.15.0 \\
\hline
\end{tabular}
\end{table}

Here, we have utilized dynamic circuits in a more clever manner. We've designed the circuits such that only those states which satisfy the column criteria proceed to construct the gates needed to check the diagonal criteria. As a result, only $N!$ states are used to build the circuit for verifying the diagonal criteria. All other states lead to the result $\ket{0}$, since no measurement is performed for those trivial states. By incorporating dynamic circuits, we significantly reduce both the time required and the complexity of the gates. This reduction stems from the elimination of time and operations spent on trivial states. This addition serves as a substantial improvement by effectively harnessing the advantages of dynamic circuits.

\section{Novel Algorithms}
\label{s5}

We have introduced two new and innovative algorithms for solving the N-Queens problem. They employ the same representation as introduced in the previous paper. We have provided comprehensive explanations for both of these algorithms in the following sections, along with relevant illustrations and examples. Sections \ref{s5a} and \ref{s5b} delve into the details of these novel algorithms, showcasing their unique approaches to addressing the N-Queens problem.

\subsection{Direct Column Algorithm}
\label{s5a}

\begin{algorithm}
\caption{DIRECT COLUMN ALGORITHM}
\label{alg:direct-column}
\begin{enumerate}
\item \textbf{Initialization of First Row:}
    \begin{enumerate}
        \item Begin by creating an N-Qubit W-State for all qubits in the first row.
    \end{enumerate}
\item \textbf{Iterative Entanglement:}
    \begin{enumerate}
        \item For each qubit in the first row ($i$-th qubit):
            \begin{enumerate}
                \item Create a controlled W-State on the second row, excluding the $i$-th qubit in the second row.
                \item The control is based on the $i$-th qubit of the first row.
                \item Similarly, for each qubit ($j$-th qubit) entangled by the W-State in the second row:
                    \begin{enumerate}
                        \item Create a W-State on the third row, except for the $i$-th and $j$-th qubits.
                        \item The control is based on the $i$-th qubit of the first row and the $j$-th qubit of the second row.
                    \end{enumerate}
                \item Continue this pattern iteratively, propagating through each row until the last row.
                \item Final Row Operation:
                    \begin{enumerate}
                        \item For the last row (with only one qubit remaining):
                        \item Perform a Toffoli gate on that qubit, controlling it with all qubits from each row that were not included in W-States in subsequent rows.
                    \end{enumerate}
            \end{enumerate}
    \end{enumerate}
\end{enumerate}
\end{algorithm}

This quantum algorithm addresses both the row criteria and column criteria concurrently. It starts by configuring the quantum circuit with combinations that satisfy both these conditions. This strategic initialization results in a substantial decrease in the search space, ultimately reaching $N!$ combinations. Remarkably, this significant reduction is achieved without the requirement for additional ancillary qubits to verify the column conditions, a necessity in previous algorithms. Following this accomplishment, the next step entails applying the same procedure for assessing the diagonal criteria, involving the utilization of ancillary qubits to check diagonals with the implementation of dynamic circuits to give only the possible configuration states. However, it's important to note an important drawback associated with this approach: as the number of input queens increases, the complexity of the quantum circuit grows exponentially. This expansion in complexity is directly linked to the quantity of input queens and can pose challenges in terms of scalability and implementation. While this approach provides a remarkable reduction in the search space and simplifies the solution process by not requiring ancillary qubits for column checks, the trade-off lies in the exponential growth of circuit complexity. This algorithm shines for smaller instances of the N-Queens problem, but its feasibility diminishes as the problem size increases due to the rapidly escalating quantum circuit complexity.

\subsection{Quantum Backtracking Algorithm}
\label{s5b}

\begin{algorithm}[H]
\caption{QUANTUM BACKTRACKING ALGORITHM}
\label{alg:quantum-backtracking-algo}

\begin{enumerate}

\item \textbf{Initialization of First Row:}
    \begin{enumerate}
        \item Begin by creating an N-Qubit W-State for all qubits in the first row.
    \end{enumerate}
\item \textbf{Iteration for Every Qubit in First Row:}
    \begin{enumerate}
        \item \textbf{Iteration over Rows:} For each qubit (i) in the first row:
            \begin{enumerate}
                \item \textbf{Disregard Certain Qubits:} Disregard qubits (i) in successive rows and those on the same diagonal as qubit (i) of the first row.
            \end{enumerate}
        \item \textbf{Create Controlled W-State:} For remaining qubits in the next row:
            \begin{enumerate}
                \item Create an N-Qubit W-State controlled by qubit (i) from the first row.
            \end{enumerate}
        \item \textbf{Iteration for Selected Qubits:} For qubits (j) in the second row on which W-State was applied:
            \begin{enumerate}
                \item \textbf{Disregard Certain Qubits:} Disregard qubits (i) and (j) in successive rows and those on the same diagonals.
            \end{enumerate}    
        \item \textbf{Create Controlled W-State:} For remaining qubits in the next row:
            \begin{enumerate}
                \item Create an N-Qubit W-State controlled by qubits (i) and (j) from the first and second rows, respectively.
            \end{enumerate}    
        \item \textbf{Repeat Iteration:} Continue iterating in a similar manner through each row until the last row.
        \item \textbf{Final Row Operation:} For the last row (with only one qubit remaining):
            \begin{enumerate}
                \item Apply a Toffoli gate on that qubit, controlling it with qubits from previous rows that weren't disregarded.
            \end{enumerate}  
    \end{enumerate}
\item \textbf{Holding Solutions}
    \begin{enumerate}
        \item For every qubit in last row, entangle them with ancillary qubit using the Controlled-NOT gate (C-NOT) with target being the ancillary qubit
    \end{enumerate}
\end{enumerate}
\end{algorithm}

The inspiration from the well-known classical backtracking algorithm is evident in the naming of this algorithm. This quantum version closely emulates the principles of the backtracking algorithm but within the context of a quantum circuit. 

\begin{figure}
    \centering
    \includegraphics[width=0.45\textwidth]{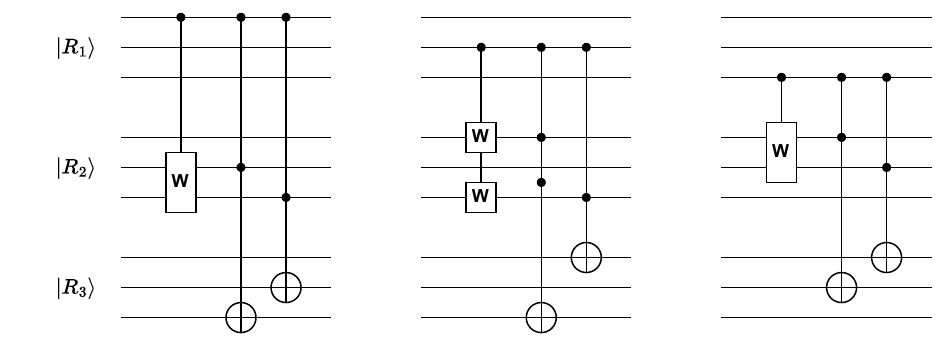}
    \caption{Example of Direct Column Algorithm using $3 \times 3$ chessboard. This depicts the steps of 2a.(\romannumeral 1) and (\romannumeral2) in Algorithm \ref{s5a}}
    \label{algo2}
\end{figure}

However, the results obtained here encompass not just a set of solutions but also non-solution states. Filtering these non-solution states requires only a single ancillary qubit. If this ancillary qubit is in the excited state $\ket{1}$, the corresponding combination is included in the solution set. If progression to the subsequent row becomes impossible due to disregarding all qubits, the algorithm employs backtracking by returning to a qubit in the same row that was previously considered and involved in creating a collective W-State. The potential solution combinations are those wherein a Toffoli gate can activate a qubit from the last row. To validate this, a single ancillary qubit is utilized, and $N$ controlled-NOT gates are applied. Each gate is controlled by a qubit from the last row. To manage the arrangement of gates, both a classical stack and a quantum stack are required. The use of classical computing resources becomes necessary for this computational aspect.

\section{Conclusion}
\label{s6}

In conclusion we have demonstrated the mapping of an N-Queens problem into a quantum computer. Improvement is done based on previous works by defining the exact structure of the state preparation circuit of W-state and the inclusion of dynamic circuits to further reduce circuit complexity. Only valid solution states are measured, while other trivial states are cancelled out by the dynamic circuit. Two novel algorithms are presented that are inspired from the present work, one being the controlled W-state and the other one based on existing techniques like the backtracking algorithm which introduces a novel approach that holds value beyond its immediate application. Although it incorporates classical computational elements, it underscores the potential to use quantum algorithms to solve the N-Queens problem. Hence future work on showing decreased time complexity for this problem using qubit-mapping is a justified prediction. Overall, this algorithm delineates a distinctive approach influenced by classical backtracking techniques, forming a bridge between classical and quantum computing realms which will pave way for the development of more efficient ways to solve the N-Queens, hence offering valuable insights for further exploration in this dynamic field.

%\section*{References}

\bibliography{sample}
\bibliographystyle{ieeetr}

% \vspace{12pt}
% \color{red}
% IEEE conference templates contain guidance text for composing and formatting conference papers. Please ensure that all template text is removed from your conference paper prior to submission to the conference. Failure to remove the template text from your paper may result in your paper not being published.

\end{document}